\documentclass[
 reprint,amsmath,amssymb,
 aps,
]{revtex4-2}

\usepackage{graphicx}
\usepackage{mhchem}
\usepackage{amsmath,amssymb}
\usepackage{xcolor}
\usepackage{siunitx}
\usepackage{float}
\usepackage{hyperref}
\usepackage{orcidlink}
\hypersetup{
    colorlinks=true,
    linkcolor=blue,
    urlcolor=blue,
    citecolor=blue
    }

\begin{document}

\title{Polarons in supersolids: Path-integral treatment of an impurity\\ in a one-dimensional dipolar supersolid}

\author{Laurent H. A. Simons \orcidlink{0009-0000-7251-5845}}\email[Contact author: ]{laurent.simons@uantwerpen.be}
\affiliation{
 Theory of Quantum and Complex Systems, Physics Department, Universiteit Antwerpen, B-2000 Antwerpen, Belgium
}

\author{Michiel Wouters \orcidlink{0000-0003-1988-4718}}
\affiliation{
 Theory of Quantum and Complex Systems, Physics Department, Universiteit Antwerpen, B-2000 Antwerpen, Belgium
}

\author{Jacques Tempere \orcidlink{0000-0001-8814-6837}}
\affiliation{
 Theory of Quantum and Complex Systems, Physics Department, Universiteit Antwerpen, B-2000 Antwerpen, Belgium
}

\date{\today}

\begin{abstract}
The supersolid phase of a dipolar Bose-Einstein condensate has an intriguing excitation spectrum displaying a band structure. Here, the dressing of an impurity in a one-dimensional dipolar supersolid with the excitations of the supersolid is studied.
The ground-state energy of the supersolid polaron is calculated using a variational path integral approach, which obtained accurate results for other polaron systems within the Bogoliubov and Fröhlich approximations. A divergence is observed at the superfluid-supersolid phase transition. The polaron radius is also computed, showing that as a function of impurity-atom interactions, the polaron can become localized to a single droplet, behaving like a small solid-state polaron.
\end{abstract}

\maketitle

\section{\label{sec:introduction}Introduction}
A supersolid is a state of matter that exhibits properties of a superfluid, such as frictionless flow, and properties of a solid, such as periodic density modulations unrelated to external potentials, at the same time \cite{tanzi2019supersolid,natale2019excitation,sohmen2021birth,bland2022two,guo2019low}. 
There are three main features associated with supersolidity: spontaneously arising periodic density modulations, global phase coherence (due to the superfluid fraction of the supersolid), and phase rigidity \cite{guo2019low}. 
Solid helium has been a candidate system for supersolidity. However, early experimental claims of supersolidity in solid helium \cite{chan2004helium} were later disproved \cite{Chan2012helium}, and no new evidence has emerged at the time of writing \cite{tanzi2019supersolid,guo2019low,leonard2017supersolid}. 
Recently, supersolidity has been observed in ultracold gases such as spin-orbit-coupled Bose-Einstein condensates (BECs) and dipolar BECs \cite{leonard2017supersolid,guo2019low,li2017stripe,tanzi2019supersolid,norcia2021two,tanzi2019observation,chomaz2019long,bottcher2019transient}.
The dipolar supersolid realization is of great interest, as supersolidity comes from the intrinsic dipole-dipole interactions and is not due to, for example, a superimposed optical lattice. Different systems and properties of dipolar supersolids have been studied recently, such as vortices, excitations, and its superfluid fraction \cite{tanzi2019supersolid,casotti2024observation,gallemi2020quantized,natale2019excitation,tanzi2021evidence,hertkorn2019fate,hertkorn2021density,hertkorn2024decoupled}. However, the polaron has yet to be studied in the context of a dipolar supersolid. The problem of an impurity in a BEC has been studied extensively for superfluid contact and dipolar Bose gases \cite{ardila2019analyzing,hu2016bose,yan2020bose,skou2022life,tempere2009feynman,grusdt2015renormalization,ardila2015impurity,vlietinck2015diagrammatic,shchadilova2016quantum,shchadilova2016polaronic,levinsen2017finite,ichmoukhamedov2019feynman,ardila2018ground,kain2014polarons,sanchez2023universal,volosniev2023non}. The polaron effect, where the impurity gets dressed by the excitations of the system, is particularly interesting as in the supersolid regime the excitation spectrum displays a band structure with two Goldstone phonon modes compared with the single phonon mode of a superfluid Bose gas \cite{natale2019excitation}.

In this paper, we study the dressing of a neutral impurity, which can have a finite magnetic moment, with the excitations of a one-dimensional (1D) dipolar supersolid. The polaronic energy is calculated within the Fröhlich and Bogoliubov approximations suited for weak impurity-atom and atom-atom interactions. 
The variational Feynman path-integral method is used, which has been shown to give results comparable with Monte Carlo methods while being numerically less intensive and time-consuming for other polaron problems \cite{vlietinck2015diagrammatic,ichmoukhamedov2019feynman,ichmoukhamedov2022general}. 
The effect of different system parameters on the polaronic energy is studied. 
A divergence in the polaronic energy is observed when crossing into the supersolid region, which can be another indication of supersolidity experimentally. 
The radius of the polaron is calculated, and it is found that for specific values, the polaron becomes localized to a single droplet of the supersolid and behaves like a small solid-state polaron, hopping from one droplet to another.

This paper is organized as follows. In Sec. II, the Hamiltonian describing the supersolid polaron is derived within the Bogoliubov and Fröhlich approximations. Section III discusses the description of the 1D dipolar supersolid itself. The wave function is calculated, and the Bogoliubov--de Gennes equation is solved to obtain the excitations. The polaronic energy and polaron radius are calculated in Sec. IV using the variational path-integral method, and the results are discussed. The conclusion and outlook are given in the last section, Sec. V.

\section{\label{sec:hamiltonian}Polaron Hamiltonian in the supersolid regime}

The Hamiltonian describing an impurity with mass $m_I$ in a 1D dipolar Bose gas is given by \cite{tempere2009feynman}
\begin{align}
    \hat{H}&=\frac{\hat{p}_I^2}{2m_I}+\sum_{q}\left(\frac{\hbar^2q^2}{2m_B}-\mu+E_t\right) \hat{a}^\dagger_{q}\hat{a}_{q} \nonumber\\
    &+ \frac{1}{2L}\sum_{k,k',q}V(q)\hat{a}^\dagger_{k'-q}\hat{a}^\dagger_{k+q}\hat{a}_{k}\hat{a}_{k'} \nonumber\\
    &+\frac{1}{L}\sum_{k,q}V_{IB}(q)\hat{\rho}_I(q)\hat{a}^\dagger_{k-q}\hat{a}_{k}.
    \label{orig}
\end{align}
The first two terms represent the free impurity and a gas of non-interacting bosons, respectively, where $\hat{a}^\dagger_{q}$ and $\hat{a}_{q}$ are the creation and annihilation operators of a dipolar atom of mass $m_B$ with momentum $q$, $\mu$ is the chemical potential, and $E_t$ is the contribution of the transverse confinement to the energy. 
The third term represents the interaction between the dipolar atoms characterized by an interaction amplitude given by the effective 1D dipolar interaction potential in momentum space $V(q)$. The interaction between the impurity and the bosons is described by the last term, where
$\hat{\rho}_I(q)=\exp(iq\hat{z}_I)$ is the impurity density and $V_{IB}(q)$ is the Fourier transform of the impurity-atom interaction potential. The 1D dipolar Bose gas is formed by tightly confining the condensate in the $x$ and $y$ directions using a harmonic trap with oscillator length $l_{\perp}$. The transverse degrees of freedom can be integrated out by using the following ansatz \cite{ilg2023ground, blakie2020variational}:
\begin{equation}
    \psi_\perp(x,y) = \frac{1}{\sqrt{\pi}\sigma l_\perp}\exp\left[-\frac{\nu x^2+y^2/\nu}{2(\sigma l_\perp)^2}\right].
\end{equation}
Here, $\sigma$ and $\nu$ are variational parameters that can be obtained by minimizing the energy of the condensate. Also, $E_t=\hbar^2(1/\nu+\nu)(1/\sigma^2+\sigma^2)/(4m_Bl_\perp)$ is the transverse trap energy corresponding to the above transverse wave function \cite{ilg2023ground, blakie2020variational}. 
There is no analytical form for the Fourier transform of the effective 1D dipolar interaction potential. Nonetheless, it has been shown that it can be described well by the following analytical expression \cite{ilg2023ground, blakie2020variational}:
\begin{equation}
   V(q)=g_{1D}^{(B)}\left[1+\epsilon^{(B)}_{dd}
   \left\{\frac{3\left[1+Qe^Q\text{Ei}(-Q)\right]}{1+\nu}-1\right\}\right], \label{Vq}
\end{equation}
with $g_{1D}^{(B)}=2\hbar^2a^{(B)}_s/(m_B\sigma^2l_\perp^2)$ the 1D $s$-wave interaction strength of the dipolar bosons and $a^{(B)}_s$ the $s$-wave scattering length of the dipolar bosons. Numerically, the infinite momentum limit is used for $V(q)$ when $Q>700$. The strength of the dipole-dipole interaction is characterized by  $\epsilon^{(B)}_{dd}=\mu_0\mu_B^2m_B/[12\pi\hbar^2a^{(B)}_s]$, where $\mu_0$ is the vacuum permeability, $\mu_B$ is the dipole moment of the atoms, $Q=\sqrt{\nu}(q\sigma l_\perp)^2/2$, and $\text{Ei}(Q)$ is the exponential integral \cite{ilg2023ground,blakie2020variational}. 

The effective 1D impurity-atom interaction potential in momentum space $V_{IB}(q_z)$ can be approximated by using the following equation \cite{blakie2020variational}:
\begin{align}
    V_{IB}(q_z)&=\frac{1}{(2\pi)^2}\int_{-\infty}^{\infty}dq_x\int_{-\infty}^{\infty}dq_y\mathcal{F}|\psi_\perp(x, y)|^2\nonumber\\&\mathcal{F}|\psi_{I,\perp}(x, y)|^2V^{(3\text{D})}_{IB}(q),
\end{align}
in addition to the following ansatz for the transverse impurity wave function $\psi_{I,\perp}(x, y)=\exp[-(x^2+y^2)/(2l_{I, \perp}^2)]/(\sqrt{\pi} l_{I, \perp})$, with $l_{I, \perp}=l_\perp\sqrt{m_B/m_I}$, assuming that the impurity feels the same transverse confinement potential as the dipolar bosons. This ansatz can break down for strong impurity-boson coupling, where transverse polaron effects become important. However, transverse effects are out of the scope of this paper. The three-dimensional (3D) impurity-atom interaction potential in momentum space $V^{(3\text{D})}_{IB}(q)$ is given by \cite{blakie2020variational,ardila2018ground}
\begin{equation}
    V^{(3\text{D})}_{IB}(q)=g^{(IB)}+g^{(IB)}\epsilon^{(IB)}_{dd}\left(3\frac{q_y^2}{q^2}-1\right).
\end{equation}
Here, $g^{(IB)}=2\pi\hbar^2a_s^{(IB)}/m_r$ is the 1D $s$-wave interaction strength of the dipolar boson-impurity interaction, $a^{(IB)}_s$ the $s$-wave scattering length for the impurity-atom interactions, $\epsilon^{(IB)}_{dd}=\mu_0\mu_I\mu_Bm_r/[6\pi\hbar^2a^{(IB)}_s]$, where $\mu_I$ is the impurity dipole moment and $m_r=(m_B^{-1}+m_I^{-1})^{-1}$ the reduced mass \cite{ardila2018ground}.

In the supersolid regime where there is a density modulation, the ground-state wave function of the dipolar gas is no longer given by the constant $\sqrt{n}$ but can be described by \cite{ilg2023ground}
\begin{equation}
    \psi_0(z)=\sqrt{n_0}\left[1+\sum_{l=1}^\infty \Delta_l\cos(lk_sz)\right],
    \label{ansatz}
\end{equation}
with $n_0=N_0/L$ the density of atoms in the zero-momentum mode, $k_s$ the momentum characterizing the density modulation, and the order parameters $\Delta_l$ describing the contribution of mode $l$ to the wave function. As translational symmetry is broken, a dipolar atom of the supersolid can be described by $q$, which is restricted to the first Brillouin zone (BZ) given by $[-k_s/2,k_s/2[$ and an integer $l$ \cite{ilg2023ground}. The creation and annihilation operators are restricted to a single BZ by including a band quantum number $l$. The momentum of the atom is then given by $q+lk_s$. This corresponds to replacing $\hat{a}^\dagger_q$ with $\hat{a}^\dagger_{q+lk_s}=\hat{a}^\dagger_{q,l}$ in Eq. (\ref{orig}). The Hamiltonian becomes
\begin{align}
    \hat{H}&=\frac{\hat{p}_I^2}{2m_I}+\sum_{q,l}\left[\frac{\hbar^2(q+lk_s)^2}{2m_B}-\mu+E_t\right]\hat{a}^\dagger_{q,l}\hat{a}_{q,l} \nonumber\\
    &+ \frac{1}{2L}\sum_{k,k',q,l,m,s}V(q+sk_s)\hat{a}^\dagger_{k'-q,m-s}\hat{a}^\dagger_{k+q,l+s}\hat{a}_{k,l}\hat{a}_{k',m} \nonumber\\
    &+\frac{1}{L}\sum_{k,q,l,l'}V_{IB}(q+lk_s)\hat{\rho}_I(q+lk_s)\hat{a}^\dagger_{k-q,l'-l}\hat{a}_{k,l'}. \label{Hpolaron}
\end{align}
Here, $\hat{a}^\dagger_{q,l}$ and $\hat{a}_{q,l}$ are the creation and annihilation operators of a dipolar atom of mass $m_B$ with momentum $q$ in mode $l$. Like the case of a neutral \cite{tempere2009feynman} and charged \cite{simons2024ion} impurity in a Bose gas, the Bogoliubov approximation and transformation can be applied. However, because of the broken translational symmetry, other modes with momentum $q=lk_s$ apart from the zero-momentum mode are also macroscopically occupied \cite{ilg2023ground}. The Bogoliubov approximation is now given by $\hat{a}_{0,l} \rightarrow \Delta_l\sqrt{N_0}/2$, with $\Delta_0 = 2$ and $\Delta_{-l}=\Delta_l$ \cite{ilg2023ground}.
A Bogoliubov transformation 
\begin{equation}
    \hat{a}_{q,l}=\sum_{l'}u^{l'}_{l}(q)\hat{\alpha}_{q,l'}+v^{l'}_{l}(q)\hat{\alpha}^\dagger_{-q,l'}
\end{equation}
can be found such that the supersolid term of the Hamiltonian [including the Lee-Huang-Yang (LHY) corrections discussed later] diagonalizes
with $u^{l'}_{l},v^{l'}_{l}$ the Bogoliubov coefficients, and $\hat{\alpha}^\dagger_{q,l},\hat{\alpha}_{q,l}$ the Bogoliubov creation and annihilation operators of an excitation with momentum $q$ and of type $l$. 
The Bogoliubov coefficients are normalized by the following relation, which is derived by enforcing the bosonic commutation relation:
\begin{equation}
    \sum_{l'}\left\{\left[u_{l}^{l'}(q)\right]^2-\left[v_{l}^{l'}(q)\right]^2\right\}=1.
\end{equation}
After applying the Bogoliubov transformation, the Hamiltonian reads
\begin{align}
    \hat{H} =& \frac{\hat{p}_I^2}{2m_I}+\sum_{q,l}\epsilon_l(q)\hat{\alpha}^\dagger_{q,l}\hat{\alpha}_{q,l}+U(\hat{z}_I)\nonumber\\&+\sum_{q,l}V_l(q,
    \hat{z}_I)\left(\hat{\alpha}_{q,l}+\hat{\alpha}^\dagger_{-q,l}\right),
    \label{frohlich}
\end{align}
where
\begin{align}
    V_l(q, \hat{z}_I) =& \frac{\sqrt{N_0}}{2L}\sum_{m,m'}V_{IB}(q+mk_s)\exp[i(q+mk_s)\hat{z}_I]\nonumber\\
    &\times \Delta_{m'}\left[u^{l}_{m-m'}(q)+v^{l}_{m-m'}(q))\right],
\end{align}
and
\begin{equation}
    U(\hat{z}_I)=\frac{n_0}{4}\sum_{l,l'}V_{IB}(lk_s)\Delta_{l'-l}\Delta_{l'}\exp[ilk_s\hat{z}_I].
\end{equation}
This is the Hamiltonian describing an impurity in a 1D dipolar supersolid within the Bogoliubov and Fröhlich approximation. 
Terms with two creation and/or annihilation operators, beyond-Fröhlich terms characterizing the absorption and emission of two Bogoliubov excitations, are neglected and out of scope for this paper. Compared with the neutral and charged Bose polaron problems, the term without any creation or annihilation operators is not constant and acts as an external periodic potential on the impurity. Therefore, it cannot be discarded and will be denoted by $U(z_I)$ from now on. 
The problem of an impurity in a dipolar supersolid has been reduced to an impurity interacting with multiple excitation bands via a Fr\"ohlich-like vertex in a periodic background potential $U(z_I)$. Here, the periodic background potential will be considered in analogy to the solid-state polaron treatment \cite{frohlich1954electrons} by introducing an interaction-dependent effective (Bloch) mass $m^*$ resulting in
\begin{align}
    \hat{H} =& \frac{\hat{p}_I^2}{2m^*}+\sum_{q,l}\epsilon_l(q)\hat{\alpha}^\dagger_{q,l}\hat{\alpha}_{q,l}\nonumber\\&+\sum_{q,l}V_l(q,
    \hat{z}_I)\left(\hat{\alpha}_{q,l}+\hat{\alpha}^\dagger_{-q,l}\right).
\end{align}
This is justified in this case, as it can be assumed that only the lowest-energy band (derived in the Appendix) contributes and the low-momentum limit can be taken, as the Bogoliubov theory does not describe the small polaron limit \cite{frohlich1954electrons,houtput2022voorbij}. The energy shift of the lowest-energy band $E_{I,0}$ is not shown in the Hamiltonian, in a similar fashion to the mean-field energy shift in other polaron problems \cite{ichmoukhamedov2019feynman}. In addition, the zone number $m$ should be put equal to zero, as only the lowest Bloch band is considered.
The order parameters will be calculated in the next section, and the Bogoliubov--de Gennes eigenproblem will be solved.

\section{\label{sec:bogoliubov} Supersolid wave function and excitations, without impurity}
To calculate the wave function and order parameters, the total energy of the supersolid needs to be minimized. The (extended) Gross-Pitaevskii (GP) energy can be used as an energy functional, which is given by \cite{ilg2023ground}
\begin{align}
    E_{\text{GP}}[\psi] &= -\int dz \psi^*(z)\frac{\hbar^2}{2m_B}\frac{d^2}{dz^2}\psi(z)+NE_t\nonumber\\&+\frac{1}{2}\int dz\int dz'V(z-z')|\psi(z)|^2|\psi(z')|^2\nonumber\\&+\frac{2}{5}\gamma\int dz |\psi(z)|^5.
\end{align}
The first line is the kinetic energy and the transverse trap energy, while the second and third lines are the dipolar interaction energy and LHY beyond-mean-field contribution, respectively. 
The parameter $\gamma$ quantifies the LHY correction and is given by \cite{ilg2023ground,schutzhold2006mean,lima2011quantum,lima2012beyond}
\begin{equation}
    \gamma=\frac{256}{15\pi}\frac{\hbar^2\left[a_s^{(B)}\right]^{5/2}}{m_B(\sigma l_\perp)^3}\int_0^1du\left[1-\epsilon_{dd}^{(B)}+3\epsilon_{dd}^{(B)} u^2\right]^{5/2}.
\end{equation}
The LHY contribution is considered, as it prevents collapse and is essential to describe the supersolid state.
Using the ansatz in Eq. (\ref{ansatz}) results in an expression for the energy in terms of the order parameters $\Delta_l$ and the variational parameters $\sigma$ and $\nu$. To obtain an analytical expression for the LHY contribution to the energy, it is assumed that the wave function is always positive or zero:
\begin{equation}
    1-\sum_{l=1}^\infty \Delta_l \geq 0,
    \label{condition}
\end{equation}
such that the absolute value can be discarded. 
This is valid close to the superfluid-supersolid transition. 
The energy divided by the number of atoms $N$ is given by
\begin{align}
    & \frac{E_{\text{GP}}[\Delta_l,k_s,\sigma,\nu]}{N}=\frac{\hbar^2k_s^2}{4m_B}\frac{\sum_{l=1}^\infty l^2\Delta_l^2}{1+\sum_{l=1}^\infty \Delta_l^2/2}+E_t \nonumber \\
    & +\frac{n}{32}\frac{\sum_{l,m,p=-\infty}^\infty\Delta_l\Delta_m\Delta_p\Delta_{l+m+p}V[(l+m)k_s]}{(1+\sum_{l=1}^\infty \Delta_l^2/2)^2} \nonumber\\
    & +\frac{\gamma n^{3/2}}{80}\frac{\sum_{l,m,p,s=-\infty}^\infty\Delta_l\Delta_m\Delta_p\Delta_s\Delta_{l+m+p+s}}{(1+\sum_{l=1}^\infty \Delta_l^2/2)^{5/2}}.
\end{align}
The above expression can be minimized numerically for $\Delta_l$, $k_s$, $\sigma$, and $\nu$, which can be substituted into Eq. (\ref{ansatz}) to obtain the ground-state wave function of the supersolid. 

The excitation spectrum and Bogoliubov coefficients can be obtained by expanding the supersolid part of the Hamiltonian up to second order in creation and annihilation operators. The LHY contribution can be considered by using the following Hamiltonian, which results in the correct speed of sound and energy correction \cite{ilg2023ground}
\begin{equation}
    \hat{H}_{\text{LHY}}=\frac{2}{5}\gamma\int dz\left[\hat{\psi}^\dagger(z)\hat{\psi}^\dagger(z)\hat{\psi}(z)\hat{\psi}(z)\right]^{5/4},
\end{equation}
with
\begin{equation}
    \hat{\psi}(z) = \psi_0(z)+\frac{1}{\sqrt{L}}\sum_{q\neq 0, l}\exp[i(q+lk_s)z]\hat{a}_{q,l}.
\end{equation}

\begin{figure}
    \centering
    \includegraphics[scale=1]{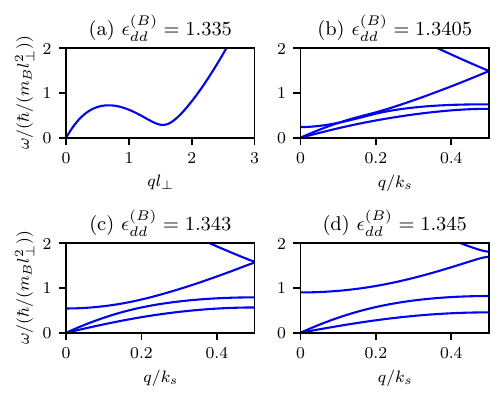}
    \caption{The excitation spectrum of a one-dimensional dipolar supersolid with parameters $n=2386.2/l_\perp$ and $a_s^{(B)}=0.005l_\perp$ for different values of $\epsilon^{(B)}_{dd}$. The critical value is given by $\epsilon^{(B)}_{dd,c}=1.34$. The results are equivalent to the ones from Refs. \cite{ilg2023ground,roccuzzo2019supersolid}.}
    \label{fig}
\end{figure}

The resulting quadratic Hamiltonian can be written as
\begin{align}
    \hat{H}_B =& \sum_{q,l,m}\left(h_{q,l,m}\hat{a}^\dagger_{q,l}\hat{a}_{q,m}+\frac{1}{2}g_{q,l,m}\hat{a}^\dagger_{q,l}\hat{a}^\dagger_{-q,m}\right.\nonumber\\
    &\left.+\frac{1}{2}g_{q,l,m}\hat{a}_{-q,m}\hat{a}_{q,l}\right),
\end{align}
with
\begin{align}
    h_{q,l,m}=&\left[\frac{\hbar^2(q+lk_s)^2}{2m_B}-\mu+E_t\right]\delta_{l,m}\nonumber\\ &+\frac{n_0}{4}\sum_sV(q+sk_s)\Delta_{m-s}\Delta_{l-s}\nonumber\\ &+\frac{n_0}{4}\sum_sV[(m-l)k_s]\Delta_{s}\Delta_{s-(l-m)}\nonumber\\ &+\frac{5}{2}\gamma n_0^{3/2}\sum_{s,s^\prime}\frac{\Delta_s\Delta_{s^\prime}\Delta_{l-m+s+s^\prime}}{8},
\end{align}
and
\begin{align}
    g_{q,l,m} = & \frac{n_0}{4}\sum_sV(q+sk_s)\Delta_{m+s}\Delta_{l-s}\nonumber\\ 
    &+\frac{3}{2}\gamma n_0^{3/2}\sum_{s,s^\prime}\frac{\Delta_s\Delta_{s^\prime}\Delta_{l+m+s+s^\prime}}{8}.
\end{align}
The Bogoliubov--de Gennes eigenproblem is then given by
\begin{equation}
    \begin{pmatrix}
h_q & g_q\\
-g_q^T & -h_{-q}^T
\end{pmatrix}\begin{pmatrix}
u^s(q)\\
v^s(-q)
\end{pmatrix}=\hbar\omega_{q,s}\begin{pmatrix}
u^s(q)\\
v^s(-q)
\end{pmatrix}.
\label{BdG}
\end{equation}
Solving the above eigenproblem results in the excitation spectrum and Bogoliubov coefficients for the dipolar supersolid.
In Fig. \ref{fig}, the excitation spectrum of a 1D dipolar supersolid is shown for realistic parameters $n = 2386.2/l_\perp$ and $a^{(B)}_s = 0.005l_\perp$ (which are taken from Ref.~\cite{ilg2023ground}) for different dipolar interaction strengths $\epsilon_{dd}^{(B)}$.  
In this case, the critical value of $\epsilon_{dd}^{(B)}$ is $\sim1.340$. For $\epsilon_{dd}^{(B)}<\epsilon_{dd,c}^{(B)}$, the excitation spectrum is the same as the analytical Bogoliubov dispersion, and a roton minimum is visible, as expected. If $\epsilon_{dd}^{(B)}$ becomes larger than the critical value, a band structure appears with two phononic modes: a crystal phonon mode and a superfluid phonon mode. The larger $\epsilon_{dd}^{(B)}$ becomes, the smaller the superfluid phonon mode becomes until the independent droplet regime is reached, where only the crystal phonon mode remains. For the values shown, the supersolid can be described by three order parameters $\Delta_1$, $\Delta_2$, and $\Delta_3$, and the condition in Eq. (\ref{condition}) is fulfilled. In addition, note that the chemical potential for the Bogoliubov--de Gennes eigenproblem is obtained analytically by using the approximation that the variational parameters are almost independent of the density (an extension of an approximation used in Ref. \cite{ilg2023ground}).

\section{\label{sec:energy}Polaronic energy and polaron radius}

For a system with action $S$, an upper bound on the free energy can be found by introducing an exactly solvable variational trial action $S_0$ and minimizing for the variational parameters. It is given by the Feynman-Jensen inequality
\begin{equation}
    F \leq F_0 + \frac{1}{\hbar\beta}\langle S-S_0\rangle_0,
\end{equation}
where $F_0$ is the free energy of the trial action and $\beta=1/(k_BT)$ the inverse temperature. 
For the trial action, a general exactly solvable quadratic trial action is used, which was also applied to other polaron problems and gave results comparable with numerically exact Monte Carlo methods (within the Fröhlich and Bogoliubov approximations) \cite{rosenfelder2001best,ichmoukhamedov2022general},
\begin{equation}
    S_0=\int \limits_0^{\hbar\beta} \frac{m^*}{2}\dot{\mathbf{r}}^2(\tau)d\tau +\frac{m^*}{2\hbar\beta}\int\limits_0^{\hbar\beta}\int\limits_0^{\hbar\beta}x(\tau-\sigma)\mathbf{r}(\tau)\cdot\mathbf{r}(\sigma)d\tau d\sigma.
\end{equation}
Here $x(\tau-\sigma)$ is a variational memory kernel \cite{rosenfelder2001best,ichmoukhamedov2022general}.
The kernel satisfies two conditions: it is imaginary-time periodic $x(\hbar\beta-\tau)=x(\tau)$ and $\int_0^{\hbar\beta} x(\tau)d\tau=0$ \cite{ichmoukhamedov2022general}. The free energy and expectation value of the trial action can be found in Ref.~\cite{ichmoukhamedov2022general}.

An imaginary-time action corresponding to the Hamiltonian in Eq. (\ref{frohlich}) can be derived like the Bose polaron problem \cite{tempere2009feynman}, and the phonon degrees of freedom can be integrated out, as it is the path integral of a forced harmonic oscillator. The resulting effective action is
\begin{align}
    S=&\int_0^{\hbar\beta}\frac{m^*}{2}\dot{z}_I(\tau)^2d\tau -\frac{1}{\hbar}\sum_{q,l}\int_0^{\hbar\beta}d\tau\int_0^{\hbar\beta}d\sigma \nonumber \\
    &\times\frac{\cosh[\omega_{q,l}(\hbar\beta/2-|\tau-\sigma|)]}{\cosh(\omega_{q,l}\hbar\beta/2)}  V^*_l[q,z_I(\tau)]V_l[q,z_I(\sigma)].
\end{align}
Using the Feynman-Jensen inequality, an upper bound is found for the free energy depending on the kernel used in the trial action $x(\nu)$. An equation for the kernel can be derived by taking the functional derivative of the free energy with respect to the kernel equal to zero. The free energy can then be calculated iteratively numerically. The set of equations are
\begin{widetext}
\begin{align}
    E_{\text{pol}} &\leq \frac{\hbar}{2\pi}\int_0^\infty d\nu \left\{\ln\left[1+\frac{x(\nu)}{\nu^2}\right]-\frac{x(\nu)}{x(\nu)+\nu^2}\right\}-\frac{n_0}{4\pi\hbar}\int_0^{k_s/2}dq \sum_{l,m',s'}V^2_{IB}(q)\Delta_{m'}\Delta_{s'}\nonumber\\ &[u^{l}_{m'}(q)+v^{l}_{m'}(q)][u^{l}_{s'}(q)+v^{l}_{s'}(q)]\int_0^\infty du \exp\left[-\omega_{q,l}u-\frac{\hbar q^2}{m^*\pi}\int_0^\infty d\nu \frac{1-\cos(\nu u)}{x(\nu)+\nu^2}\right],
\end{align}
and 
\begin{align}
    x(\nu) &= \frac{n_0}{\pi\hbar m^*}\int_0^{k_s/2}dq \sum_{l,m',s'}q^2V^2_{IB}(q)\Delta_{m'}\Delta_{s'}[u^{l}_{m'}(q)+v^{l}_{m'}(q)][u^{l}_{s'}(q)+v^{l}_{s'}(q)]\nonumber\\ &\int_0^\infty du \sin^2(\nu u/2)\exp\left[-\omega_{q,l}u-\frac{\hbar q^2}{m^*\pi}\int_0^\infty d\nu \frac{1-\cos(\nu u)}{x(\nu)+\nu^2}\right].
\end{align}
\end{widetext}
The iterative procedure is started using the Lee-Low-Pines kernel given by $x(\nu)=0$. 
We consider the upper bound to be converged if the difference with the last value is $\le0.1\%$.
A Gauss-Legendre quadrature with $20$ points with $100$ subintervals was used as in Refs. \cite{rosenfelder2001best, simons2024ion} for the polaronic energy calculation. Another method, perturbation theory, is used to compare the polaronic energy results with the path-integral result. The equation for the polaronic energy using perturbation theory is given by
\begin{align}
    E_{\text{pert}} &= -\frac{n_0}{4\pi\hbar}\sum_{l,s,p}\Delta_{s}\Delta_{p}\int_0^{k_s/2}dq V^2_{IB}(q)\nonumber\\&\times\frac{[u^l_{s}(q)+v^l_{s}(q)][u^l_{p}(q)+v^l_{p}(q)
    }{\omega_{q,l}+\frac{\hbar q^2}{2m^*}}.
\end{align}
This is equivalent to making the kernel equal to zero in our path-integral polaronic energy equation. In addition, the constant energy shift of the lowest impurity band in the periodic potential $E_{I,0}$ is not considered in the definition of the polaronic energy as it can be interpreted similar to the mean-field energy shift in other polaron problems \cite{ichmoukhamedov2019feynman}.

\begin{figure}[H]
    \centering
    \includegraphics[scale=1]{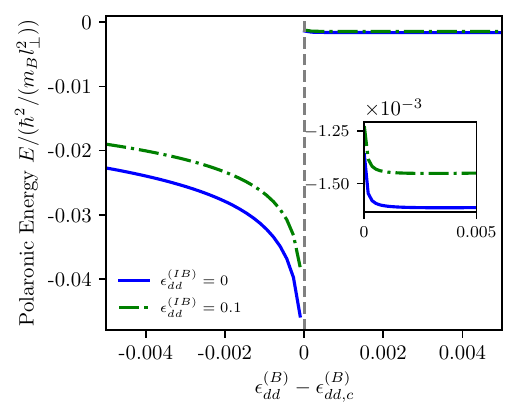}
    \caption{The polaronic energy for $n = 2386.2/l_\perp$, $a^{(B)}_s = 0.005l_\perp$, $a^{(IB)}_s=0.004l_\perp$, and $m_I=m_B$ as a function of $\epsilon^{(B)}_{dd}-\epsilon^{(B)}_{dd,c}$. The blue line is the polaronic energy for the non-dipolar impurity, while the green dash-dotted line is the polaronic energy of an impurity with a finite magnetic moment. The gray dashed line indicates the transition from the superfluid to the supersolid phase. The inset shows the variation of the supersolid energy more clearly.}
    \label{fig2}
\end{figure}

\begin{figure}[H]
    \centering
    \includegraphics[scale=1]{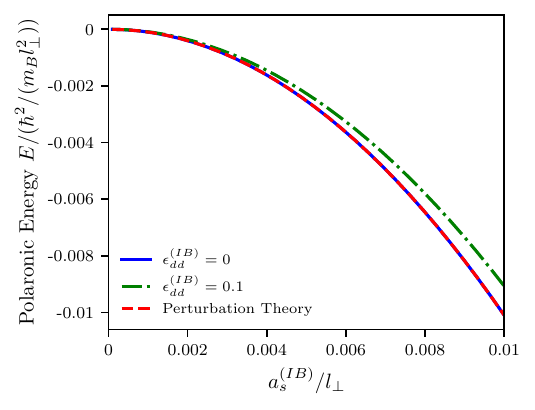}
    \caption{The polaronic energy as a function of the s-wave scattering length $a^{(IB)}_s$ in the supersolid regime for $n = 2386.2/l_\perp$, $a^{(B)}_s = 0.005l_\perp$, $m_I=m_B$, $\epsilon^{(B)}_{dd}=1.345$, and $\epsilon^{(IB)}_{dd}=0$ (blue line) or $0.1$ (green dash-dotted line). The red dashed line shows the perturbation theory result for the non-dipolar impurity which agrees very well with the path-integral result.}
    \label{fig3}
\end{figure}

The polaronic energy $E_{\text{pol}}=F-E_{I,0}$, with $E_{I,0}=nV_{IB}(0)$ in the superfluid regime, for $n = 2386.2/l_\perp$, $a^{(B)}_s = 0.005l_\perp$, $a^{(IB)}_s=0.004l_\perp$, and $m_I=m_B$ across the transition as a function of $\epsilon^{(B)}_{dd}-\epsilon^{(B)}_{dd,c}$ is shown in Fig. \ref{fig2}. 
The blue line is for $\epsilon^{(IB)}_{dd}=0$, where the impurity does not have a magnetic moment and is not dipolar. 
In the superfluid regime, the energy decreases as the system gets closer to the phase transition. There is a divergence visible at the transition $\epsilon^{(B)}_{dd}=\epsilon^{(B)}_{dd,c}$, which was also seen in Ref. \cite{ardila2018ground} using perturbation theory in the superfluid regime for a two-dimensional dipolar gas. 
The divergence from the superfluid regime is driven by the softening of the roton mode \cite{ardila2018ground}. The divergence will persist even when beyond-Fr\"ohlich corrections are added as the impurity-atom scattering length used here is small. However, the divergence of the number of excitations dressing the impurity will result in a large local depletion where Bogoliubov theory used here will fail.
In the supersolid regime, the energy decreases as the dipolar interaction strength becomes larger. The green dash-dotted line shows the polaronic energy when the impurity is dipolar and has a finite magnetic moment resulting in $\epsilon^{(IB)}_{dd}=0.1$. 
The additional dipole-dipole interaction for the impurity decreases the impurity-atom interaction potential, resulting in an increase of the polaronic energy and a weaker polaron. In Fig. \ref{fig3}, the dependence of the polaronic energy on the $s$-wave scattering length $a^{(IB)}_s$ in the supersolid regime is shown for $n = 2386.2/l_\perp$, $a^{(B)}_s = 0.005l_\perp$, $m_I=m_B$, $\epsilon^{(B)}_{dd}=1.345$, and $\epsilon^{(IB)}_{dd}=0$ or $0.1$. 
Like the charged Bose polaron problem \cite{simons2024ion}, the energy is always negative and decreases as $a^{(IB)}_s$ increases. 
Even when the impurity has a finite magnetic moment, which increases the polaronic energy, the polaronic energy is always negative within the Fr\"ohlich approximation used here. The energy obtained via perturbation theory is also plotted. As can be seen, the difference between perturbation theory and the path-integral approach is small.
In the supersolid regime, it is possible that, for strong interaction strengths, the polaron becomes localized to a single droplet (or single lattice point) of the supersolid. 

\begin{figure}[H]
    \centering
    \includegraphics[scale=1]{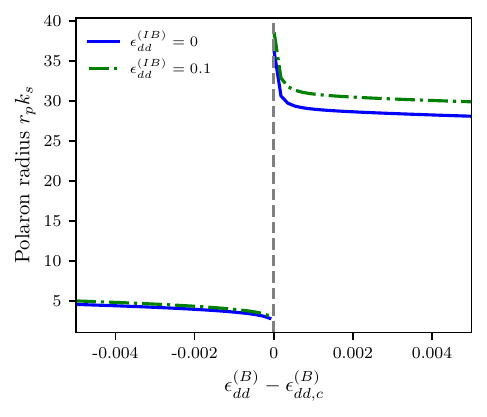}
    \caption{The polaron radius $r_p$ in units of $1/k_s$ as a function of $\epsilon^{(B)}_{dd}-\epsilon^{(B)}_{dd,c}$ for $n = 2386.2/l_\perp$, $a^{(B)}_s = 0.005l_\perp$, $a^{(IB)}_s=0.004l_\perp$, and $m_I=m_B$. The blue line is the polaronic energy for the non-dipolar impurity, while the green dash-dotted line is the polaronic energy of an impurity with a finite magnetic moment. The gray dashed line indicates the transition from the superfluid to the supersolid phase.}
    \label{fig4}
\end{figure}

\begin{figure}[H]
    \centering
    \includegraphics[scale=1]{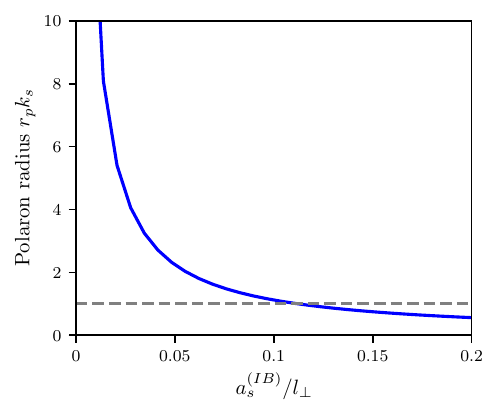}
    \caption{The polaron radius $r_p$ in units of $1/k_s$ in the supersolid regime as a function of $a^{(IB)}_s$ for $n = 2386.2/l_\perp$, $a^{(B)}_s = 0.005l_\perp$, $m_I=m_B$, $\epsilon^{(B)}_{dd}=1.345$, and $\epsilon^{(IB)}_{dd}=0$. The gray line indicates the $r_p=1/k_s$ threshold where the impurity gets localized to a single droplet of the supersolid.}
    \label{fig5}
\end{figure}

To study the localization of the polaron, the polaron radius needs to be calculated. The polaron radius can be calculated using \cite{houtput2022voorbij,simons2024ion}
\begin{equation}
    r_p=\sqrt{\frac{\pi\hbar}{4m^*}\left[\int_0^\infty d\nu \frac{x(\nu)}{x(\nu)+\nu^2}\right]^{-1}}.
\end{equation}
The above equation was used to calculate the polaron radius for the charged Bose polaron \cite{simons2024ion} and the anharmonic solid-state polaron \cite{houtput2022voorbij}. 
Other methods are also possible \cite{tempere2009feynman,mitra1987polarons}. 
In Fig. \ref{fig4}, the polaron radius multiplied by the density modulation momentum $k_s$ is plotted across the transition for $n = 2386.2/l_\perp$, $a^{(B)}_s = 0.005l_\perp$, $a^{(IB)}_s=0.004l_\perp$, and $m_I=m_B$ against $\epsilon^{(B)}_{dd}-\epsilon^{(B)}_{dd,c}$. 
The blue line is for a nondipolar impurity $\epsilon^{(IB)}_{dd}=0$, while the green dash-dotted line is for an impurity with a finite magnetic moment $\epsilon^{(IB)}_{dd}=0.1$. The polaron radius behaves like the ground-state energy, as a divergence is observed at the transition due to softening of the roton mode.
The finite magnetic moment of the impurity increases the size of the polaron. For $k_s$ in the superfluid regime, the value of $k_s$ for the supersolid close to the transition is used.
For the values used here, $r_pk_s>1$, indicating a polaron that is not localized to a droplet. 
Figure \ref{fig5} shows the polaron radius as a function of $a^{(IB)}_s$ for $n = 2386.2/l_\perp$, $a^{(B)}_s = 0.005l_\perp$, $m_I=m_B$, $\epsilon^{(B)}_{dd}=1.345$, and $\epsilon^{(IB)}_{dd}=0$. 
The polaron radius decreases as $a^{(IB)}_s$ increases, as expected, and at a certain value of $a^{(IB)}_s$, the polaron radius falls below $1/k_s$, and the polaron is localized to a single droplet. 
In this case, polaron motion will be best described by hopping from one droplet to another, like a small solid-state polaron. In Fig. \ref{appendixfig}, where the Bloch mass is plotted, it is also visible that, in the regime that the polaron becomes localized, the Bloch mass is very large. Compared with a polaron in an optical lattice, the localization comes from intrinsic interactions with the condensate.
Note that the localization of the polaron happens in the intermediate- or strong-coupling regime where beyond-Fr\"ohlich corrections probably have a substantial effect. In addition, the Bloch mass is large for the values used here, and the other higher Bloch bands are not considered, which can affect the result. An exact incorporation of the periodic background potential is out of the scope of this paper and left for future investigation.

\section{\label{sec:conclusion}Conclusion and Outlook}
The wave function of a 1D dipolar supersolid and the corresponding excitations have been reviewed and discussed.
The ground-state energy of an impurity dressed by the excitations of the supersolid has been calculated using the Feynman path-integral method within the Bogoliubov and Fröhlich approximations. 
A divergence was observed when the supersolid regime was entered. 
An additional impurity-atom dipole interaction from a finite impurity magnetic moment results in a larger ground-state energy. 
The polaron radius has also been calculated, showing that, for most values, the polaron radius is larger than the size of a single droplet of the supersolid. 
However, for large enough impurity-atom interaction strengths, the polaron can get localized to a single droplet, behaving like a small polaron. 
An interesting next step would be to experimentally study the system of an impurity in a dipolar supersolid and compare the results with the path-integral method shown here. 
Additionally, there are no numerically exact Monte Carlo results yet for comparison.

\begin{acknowledgments}
We gratefully acknowledge fruitful discussions with T.~Bland and T.~Ichmoukhamedov. We acknowledge financial support by the Research Foundation - Flanders (FWO), Projects No. GOH1122N, No. G061820N, No. G060820N, No. G0AIY25N, and No. G0A9F25N, and by the University Research Fund (BOF) of the University of Antwerp.
\end{acknowledgments}

\appendix*
\section{Impurity Bloch mass}

\begin{figure*}
    \centering
    \includegraphics[scale=1]{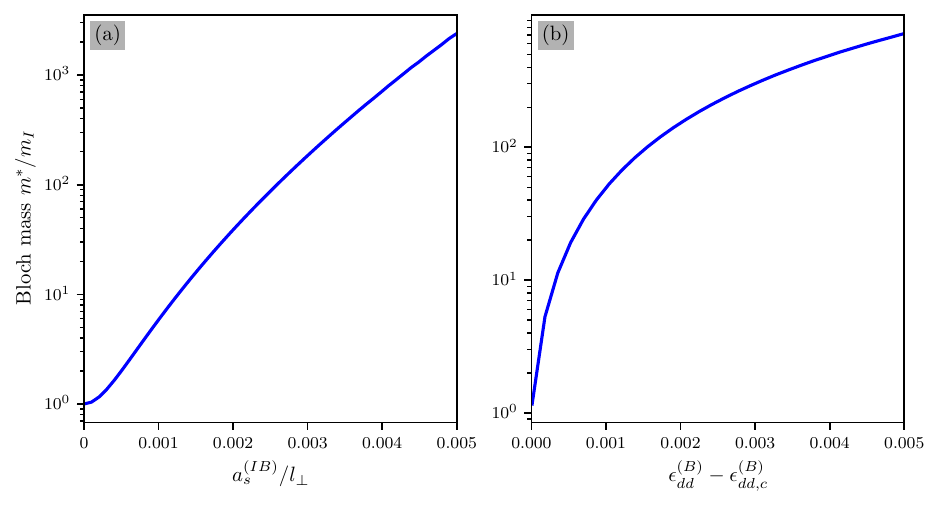}
    \caption{Impurity Bloch mass $m^*/m_I$ as a function of (a) the $s$-wave impurity-boson scattering length $a_s^{(IB)}$ for $\epsilon_{dd}^{(B)}=1.345$, $n=2386.2/l_\perp$, $a_s^{(B)}=0.005l_\perp$, $m_I=m_B$, and $\epsilon^{(IB)}_{dd}=0$ and (b) the relative dipolar interaction strength $\epsilon_{dd}^{(B)}-\epsilon_{dd,c}^{(B)}$ for $a_s^{(IB)}=0.004l_\perp$, $n=2386.2/l_\perp$, $a_s^{(B)}=0.005l_\perp$, $m_I=m_B$, and $\epsilon^{(IB)}_{dd}=0$ on a log scale.}
    \label{appendixfig}
\end{figure*}

The effective impurity mass due to the periodic potential $U(z_I)$ can be derived by solving the Schr\"odinger equation using the Bloch theorem \cite{Ashcroft}
\begin{equation}
    \left[\frac{\hbar^2}{2m_I}(k-lk_s)^2-E_n\right]C_{k-lk_s}+\sum_{m}U_{m-l}C_{k-mk_s}=0,
\end{equation}
with $\psi_I(z)=\sum_{k}C_k\exp(ikz)$ and $U_l=n_0V_{IB}(lk_s)\sum_{l'}\Delta_{l'-l}\Delta_{l'}/4$.
The above matrix equation can be solved, and the energy bands $E_n$ can be numerically calculated. For most values used in this paper, the lowest-energy band is extremely flat, resulting in a very high effective mass $m^*=[\partial^2 E/\partial^2 (\hbar k)]_{k=0}^{-1}$. The dependence of the effective mass $m^*$ on the impurity-boson $s$-wave scattering length $a_s^{(IB)}$ and the relative dipolar interaction strength $\epsilon_{dd}^{(B)}$ is shown in Fig. \ref{appendixfig}. It can be seen that it increases as a function of both $a_s^{(IB)}$ and $\epsilon_{dd}^{(B)}$. The closer to the transition, the smaller the density modulation of the condensate and periodic potential, resulting in a smaller effective mass which reaches $m_I$ at the transition.

\end{document}